\documentclass{ws-ijmpcs}

\begin{document}

\markboth{Di Sciascio}
{Instructions for Typing Manuscripts (Paper's Title)}

%
\catchline{}{}{}{}{}
%

\title{MEASUREMENT OF COSMIC RAY SPECTRUM AND ANISOTROPY WITH ARGO-YBJ}

\author{G. DI SCIASCIO on behalf of the ARGO-YBJ Collaboration}

\address{INFN, Sezione di Roma Tor Vergata, Viale della Ricerca Scientifica 1, Roma, Italy I-00133.\\
giuseppe.disciascio@roma2.infn.it}

\maketitle

\begin{history}
\received{Day Month Year}
\revised{Day Month Year}
\end{history}

\begin{abstract}

In this paper we report on the observation of the anisotropy of cosmic ray arrival direction at different angular scales with ARGO-YBJ. Evidence of new few-degree excesses throughout the sky region 195$^{\circ}\leq$ R.A. $\leq$ 315$^{\circ}$ is presented for the first time.

We report also on the measurement of the light-component (p+He) spectrum of primary cosmic rays in the range 5 - 200 TeV. 

\end{abstract}

\ccode{PACS numbers: 96.50.S-, 96.50.sd, 96.50.sb}

\section{Introduction}	

The measurement of the anisotropies of cosmic rays (CRs) arrival direction at different angular scales is complementary to the study of their energy spectrum and chemical composition, to understand their origin and propagation.

As CRs are mostly charged nuclei, their arrival direction is deflected and highly isotropized by the action of galactic magnetic field (GMF) they propagate through before reaching the Earth atmosphere. The GMF is the superposition of regular field lines and chaotic contributions. Altough the strength of the non-regular component is still under debate, the local total intensity is supposed to be $B=2\div 4\textrm{ $\mu$G}$. In such a field, the gyroradius of CRs is given by $r_{a.u.}=100\,R_{\textrm{\scriptsize{TV}}}$,
where $r_{a.u.}$ is in astronomic units and $R_{\textrm{\scriptsize{TV}}}$ is in TeraVolt.
However, different experiments \cite{nagashima,kam07,tibet06,milagro09,eastop09,icecube11} observed an energy-dependent \emph{"large scale"} anisotropy in the sidereal time frame with an amplitude of about 10$^{-4}$ - 10$^{-3}$, suggesting the existence of two distint broad regions, one showing an excess of CRs (called "tail-in"), distributed around 40$^{\circ}$ to 90$^{\circ}$ in Right Ascension (R.A.). The other a deficit (the so-called "loss cone"), distributed around 150$^{\circ}$ to 240$^{\circ}$ in R.A..

In the last years the Tibet AS$\gamma$ \cite{amenomori07} and Milagro \cite{milagro08} Collaborations reported evidence of the existence of a medium angular scale anisotropy contained in the tail-in region.
The observation of similar small scale anisotropies has been recently claimed by the Icecube experiment in the Southern hemisphere \cite{icecube11}.
So far, no theory of CRs in the Galaxy exists which is able to explain the origin of these different anisotropies leaving the standard model of CRs and that of the local GMF unchanged at the same time.
A joint analysis of concurrent data recorded by different experiments in both hemispheres, as well as a correlation with other observables like the interstellar energetic neutral atoms distribution \cite{ibex09}, should be a high priority to clarify the observations.

The ARGO-YBJ experiment, located at the YangBaJing Cosmic Ray Laboratory (Tibet, P.R. China, 4300 m a.s.l., 606 g/cm$^2$), is an air shower array able to detect the cosmic radiation with an energy threshold of a few hundred GeV. 
The full detector is in stable data taking since November 2007 with a duty cycle larger than 85\%. The trigger rate is 3.6 kHz. The detector characteristics and performance are described in \cite{moon11}. The low energy threshold and the high duty cycle make ARGO-YBJ suitable to overlap direct measurements of the CR primary spectrum in a wide region well below 100 TeV, not accessible by other EAS experiments. 

In this paper the observation of CR anisotropy at different angular scales with ARGO-YBJ is reported as a function of the primary energy.
We report also on the measurement of the differential energy spectrum of the primary CR light component (p+He) in the energy region (5 - 200) TeV.

\section{CR anisotropy}

To study the anisotropy at different angular scales the isotropic background of CRs has been estimated with two well-known methods: the equi-zenith angle method \cite{amenomori05} and the direct integration method \cite{Fleysher}.
The equi-zenith angle method, used to study the large scale anisotropy, is able to eliminate various spurious effects caused by instrumental and environmental variations, such as changes in pressure and temperature that are hard to control and tend to introduce systematic errors in the measurement.
The direct integration method, based on time-average, relies on the assumption that the local distribution of the incoming CRs is slowly varying and the time-averaged signal may be used as a good estimation of the background content. 
Time-averaging methods act effectively as a high-pass filter, not allowing to inspect features larger than the time over which the background is computed (i.e., 15$^{\circ}$/hour$\times \Delta t$ in R.A.). The time interval used to compute the average spans $\Delta t$= 3 hours and makes us confident the results are reliable for structures up to $\approx$35$^{\circ}$ wide. 
%
\begin{figure}[t!]
  \hspace{-0.8cm}
\begin{minipage}[t]{.5\linewidth}
\begin{center}
 \includegraphics[width=0.85\textwidth]{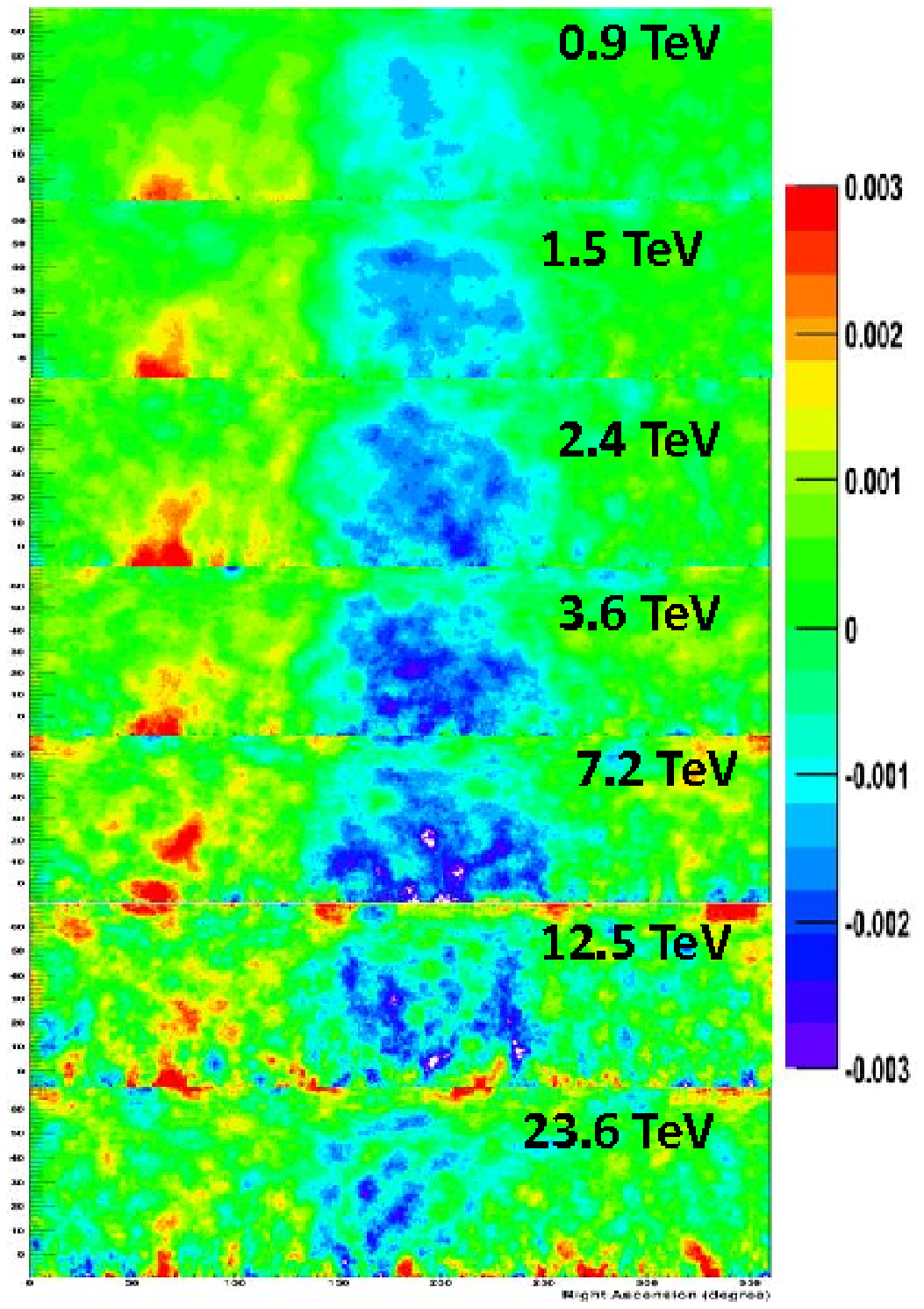}\\
   \end{center}
\end{minipage}\hfill
\begin{minipage}[t]{.52\linewidth}
  \begin{center}
  \includegraphics[width=0.95\textwidth]{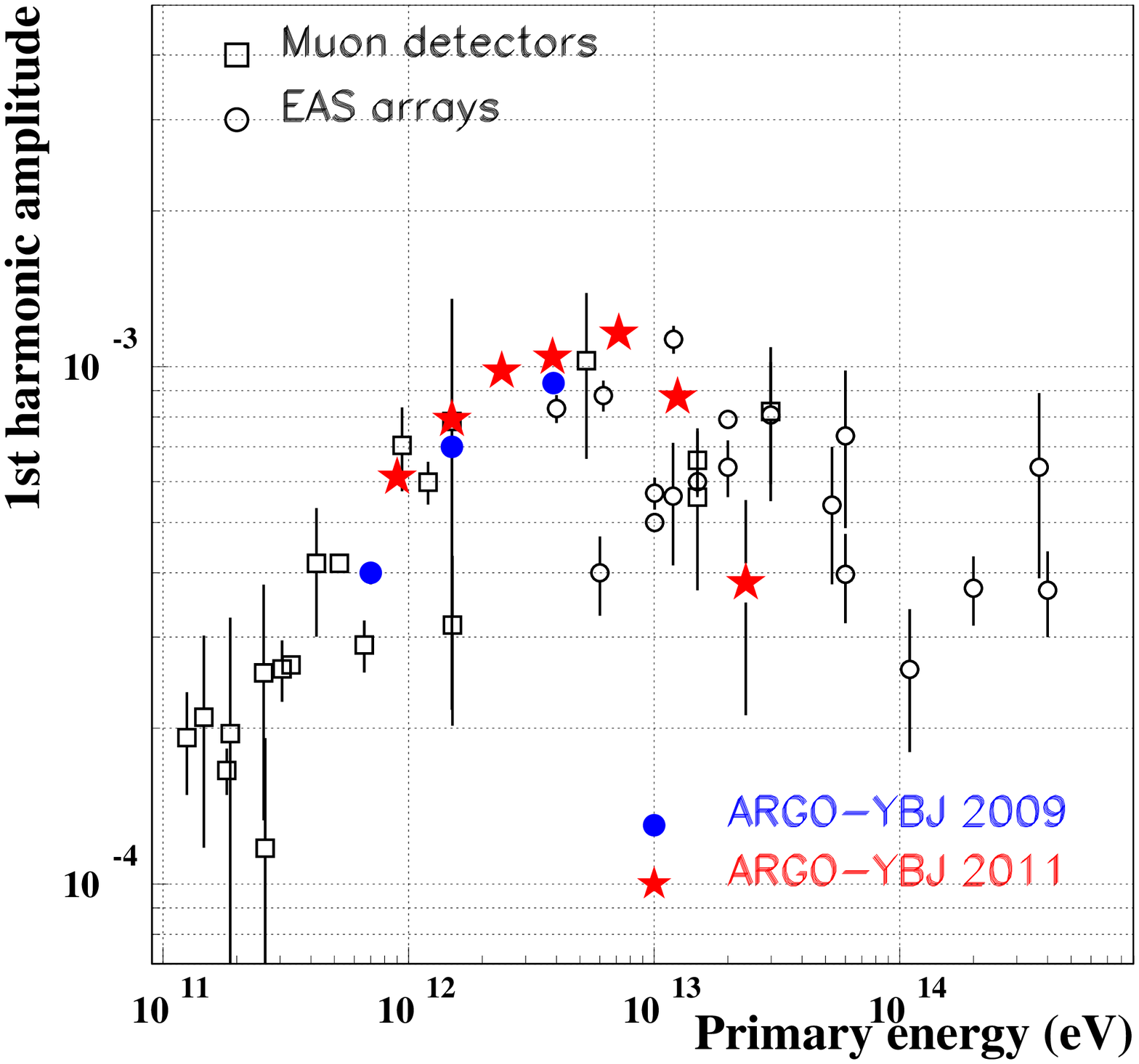}\\
  \end{center}
\end{minipage}\hfill
  \caption{Left plot: Large scale CR anisotropy observed by ARGO-YBJ as a function of the energy. The color scale gives the relative CR intensity.
  Right plot: Amplitude of the first harmonic as a function of the energy, compared to other measurements.} 
\label{fig1}
\end{figure}

\subsection{Large Scale Anisotropy}
The observation of the CR large scale anisotropy by ARGO-YBJ is shown in the left plot of Fig. \ref{fig1} as a function of the primary energy up to about 25 TeV. 
The data used in this analysis were collected by ARGO-YBJ from 2008 January
to 2009 December with a reconstructed zenith angle $\leq$ 45$^{\circ}$.
The so-called \textit{`tail-in'} and \textit{`loss-cone'} regions, correlated to an enhancement and a deficit of CRs, are clearly visible with a statistical significance greater than 20 s.d..
The tail-in broad structure appears to dissolve to smaller angular scale spots with increasing energy.
To quantify the scale of the anisotropy we studied the 1-D R.A. projections integrating the sky maps inside a declination band given by the field of view of the detector. Therefore, we fitted the R.A. profiles with the first two harmonics. The resulting amplitude of the first harmonic is plotted in the right plot of Fig. \ref{fig1} where is compared to other measurements as a function of the energy. The ARGO-YBJ results are in agreement with other experiments suggesting a decrease of the anisotropy first harmonic amplitude with increasing energy above 10 TeV.
%
\begin{figure}[t!]
  \hspace{-0.8cm}
\begin{minipage}[t]{.5\linewidth}
\begin{center}
 \includegraphics[width=0.95\textwidth]{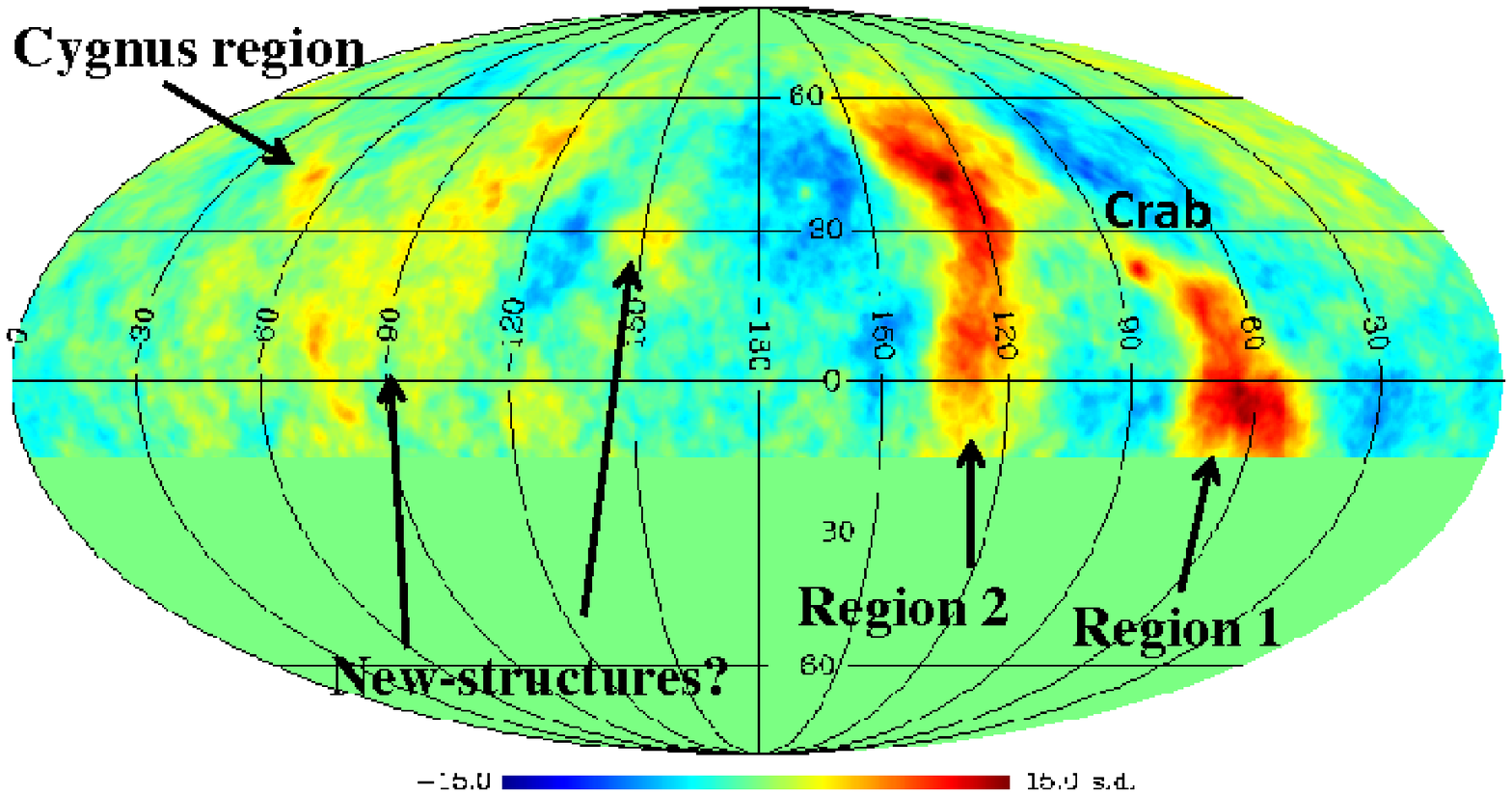}\\
   \end{center}
\end{minipage}\hfill
\begin{minipage}[t]{.52\linewidth}
  \begin{center}
  \includegraphics[width=0.85\textwidth]{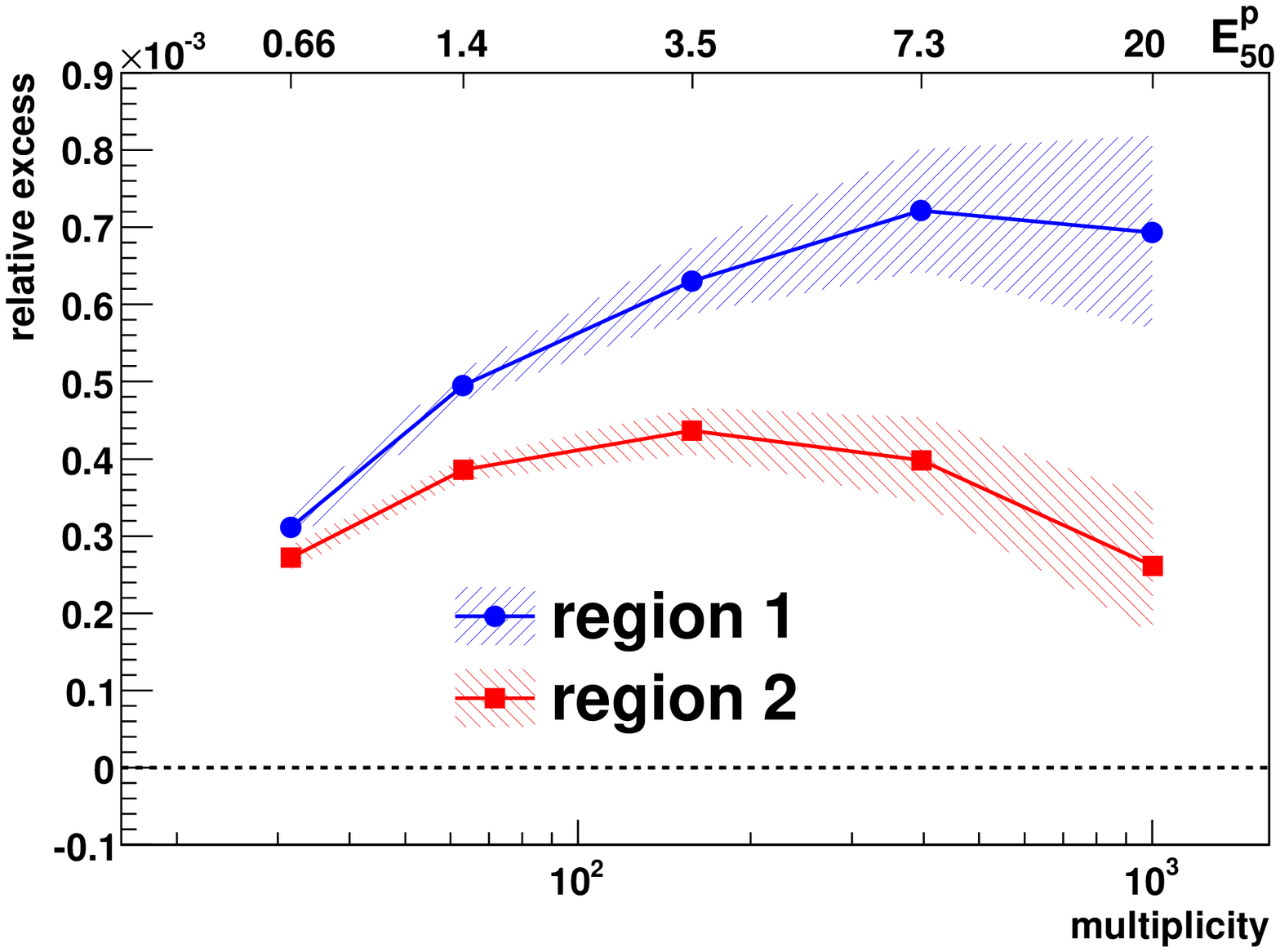}\\
  \end{center}
\end{minipage}\hfill
  \caption{Left plot: Medium scale CR anisotropy observed by ARGO-YBJ. The color scale gives the statistical significance of the observation in standard deviations. 
  Right plot: Size spectrum of the regions 1 and 2. The vertical axis represents the relative excess (Ev-Bg)/Bg. The upper scale shows the corresponding proton median energy. The shadowed areas represent the 1$\sigma$ error band.} 
\label{fig2}
\end{figure}
%

\subsection{Medium Scale Anisotropy}
The left plot of the Fig. \ref{fig2} shows the ARGO-YBJ sky map in equatorial coordinates.
The analysis refers to events collected from November 2007 to May 2011 after the following selections: (1) $\geq$25 shower particles on the detector; (2) zenith angle of the reconstructed showers $\leq$50$^{\circ}$.
The triggering showers that passed the selection were about 2$\cdot$10$^{11}$. The zenith cut selects the declination region $\delta\sim$ -20$^{\circ}\div$ 80$^{\circ}$.
According to the simulation, the median energy of the isotropic cosmic ray proton flux is E$_p^{50}\approx$1.8 TeV (mode energy $\approx$0.7 TeV).

The most evident features are observed by ARGO-YBJ around the positions $\alpha\sim$ 120$^{\circ}$, $\delta\sim$ 40$^{\circ}$ and $\alpha\sim$ 60$^{\circ}$, $\delta\sim$ -5$^{\circ}$, positionally coincident with the regions detected by Milagro \cite{milagro08}. These regions, named ``region 1'' and ``region 2'', are observed with a statistical significance of about 14 s.d.. 
The deficit regions parallel to the excesses are due to a known effect of the analysis, that uses also the excess events to evaluate the background, overestimating the background.
On the left side of the sky map, several possible new extended features are visible, though less intense than those aforementioned. The area $195^{\circ}\leq R.A. \leq 315^{\circ}$ seems to be full of few-degree excesses not compatible with random fluctuations (the statistical significance is more than 6 s.d. post-trial). 
The observation of these structures is reported here for the first time and together with that of regions 1 and 2 it may open the way to an interesting study of the TeV CR sky.
To figure out the energy spectrum of the excesses, data have been divided into five independent shower multiplicity sets. The number of events collected within each region are computed for the event map (Ev) as well as for the background one (Bg). The relative excess (Ev-Bg)/Bg is computed for each multiplicity interval. 
The result is shown in the right plot of the Fig. \ref{fig2}. Region 1 seems to have spectrum harder than isotropic CRs and a cutoff around 600 shower particles (proton median energy E$^{50}_p$ = 8 TeV). On the other hand, the excess hosted in region 2 is less intense and seems to have a spectrum more similar to that of isotropic cosmic rays.
The steepening from 100 shower particles on (E$_p^{50}$ = 2 TeV) is
likely related to efficiency effects. 
We note that, in order to filter the global anisotropy, we used a method similar to the one used by Milagro and Icecube. 
Further studies using different approaches are on the way.

\section{Measurement of the light component spectrum of CRs}

\begin{figure}[t!]
\begin{center}
\epsfysize=5.5cm \hspace{0.5cm}
\epsfbox{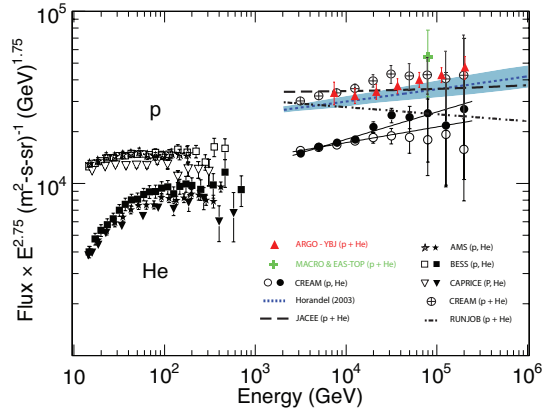} 
\caption[h]{Light-component (p+He) spectrum of primary CRs measured by ARGO-YBJ compared with other experimental results.}
\label{fig:light_spectrum}
  \end{center}
\end{figure}

Requiring quasi-vertical showers ($\theta<$30$^{\circ}$) and applying a selection criterion based on the particle density, a sample of events mainly induced by protons and helium nuclei with the shower core inside a fiducial area (with radius $\sim$28 m) has been selected. The contamination by heavier nuclei is found negligible. An unfolding technique based on the Bayesian approach has been applied to the strip multiplicity distribution in order to obtain the differential energy spectrum of the light-component (protons and helium nuclei) in the energy range (5 - 200) TeV. 
The spectrum measured by ARGO-YBJ is compared with other experiments in Fig. \ref{fig:light_spectrum}.  
Systematic effects due to different hadronic models and to the selection criteria do not exceed 10\%.
The ARGO-YBJ data agree remarkably well with the values obtained by adding up the proton and helium fluxes measured by CREAM both concerning the total intensities and the spectrum \cite{cream11}. The value of the spectral index of the power-law fit representing the ARGO-YBJ data is -2.61$\pm$0.04, which should be compared to $\gamma_p$ = -2.66$\pm$0.02 and $\gamma_{He}$ = -2.58$\pm$0.02 obtained by CREAM.
The present analysis does not allow the determination of the individual proton and Helium contribution to the measured flux, but the ARGO-YBJ data clearly exclude the RUNJOB results \cite{runjob}. 
We emphasize that for the first time direct and ground-based measurements overlap for a wide energy range thus making possible the cross-calibration of the different experimental techniques.

\end{document}